\documentclass[a4paper]{iopart}

\usepackage{mathrsfs}
\usepackage{iopams}

\usepackage{psfrag}
\usepackage{graphicx}
\usepackage{dsfont}
\usepackage[cdot,textstyle, amssymb]{SIunits}
\usepackage{subfigure,multirow,rotating}
\newcommand{\num}[1]{#1}
\newcommand{\SI}[2]{\num{#1}~{#2}}

\newcommand{\tensorp}{\otimes}
\newcommand{\identity}{\mathds{1}}
\newcommand{\p}[1]{\!\left( #1 \right)}

\renewcommand{\d}[1]{\!\mathrm{d}{#1}\ \!}
\newcommand{\abs}[1]{\left| #1\right| }
\newcommand{\T}{\mathscr{T}}  
\renewcommand{\vec}[1]{\boldsymbol{#1}}

\newcommand{\symInput}{\textrm{input}}

\newenvironment{align}{\begin{eqnarray}}{\end{eqnarray}}
\newenvironment{align*}{\begin{eqnarray}}{\end{eqnarray}}

\newcommand{\ket}[1]{| #1 \rangle}
\newcommand{\bra}[1]{\langle #1 |}
\newcommand{\braket}[1]{\langle #1 \rangle}

\begin{document}
\title{Quantum learning by measurement and feedback}
\author{S. Gammelmark and K. M{\o}lmer}
\address{Lundbeck Foundation Theoretical Center for Quantum System
Research, Department of Physics and Astronomy, University of Aarhus, DK 8000 Aarhus C, Denmark}
\ead{\mailto{gammelmark@phys.au.dk}, \mailto{moelmer@phys.au.dk}}
\date{Unknown}
\pacs{03.67.Ac, 07.05.Mh, 87.19.lr}

\begin{abstract}
  We investigate an approach to quantum computing in which quantum
  gate strengths are parametrized by quantum degrees of freedom. The
  capability of the quantum computer to perform desired tasks is
  monitored by measurements of the output and gradually improved by
  successive feedback modifications of the coupling strength
  parameters. Our proposal only uses information available in an
  experimental implementation, and is demonstrated with simulations on
  search and factoring algorithms.
\end{abstract}


\section{Introduction}

Quantum information science deals with the use of quantum resources to
speed up quantum computing, and to enable new features in quantum
communication \cite{NielsenChuang}. The usual paradigm of quantum
computing consists of well defined physical storage modes of
individual qubits and the existence of a universal set of basic gate
operations, from which any unitary operation can be constructed on a
full quantum register. The implementation of the universal set of
gates in different physical proposals constitutes the back bone for
most experimental work on quantum computing. Proof of the ability to
perform any computation does not necessarily provide an efficient
encoding of the computation in terms of universal gates, and it does
not in any simple manner point to optimal performance, e.g., under
restrictions set by physically motivated cost functions.

\begin{figure}
  \centering
  \subfigure[~]{
  \begin{psfrags}
    \includegraphics[width=0.4\textwidth]{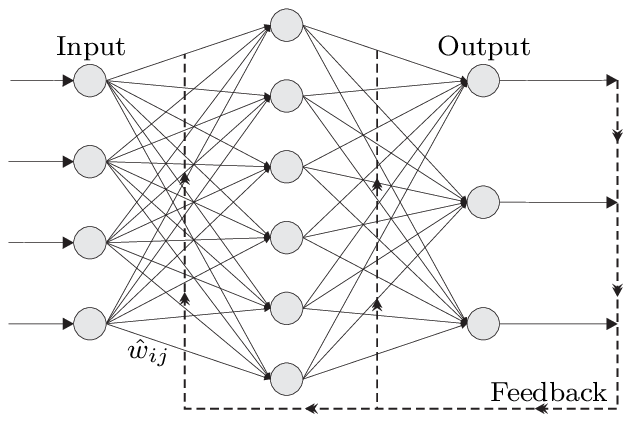}
    \label{fig:modelNN}
  \end{psfrags}
  } \hspace{0.05\textwidth}  \subfigure[~]{
  \begin{psfrags}
    \includegraphics[width=0.4\textwidth]{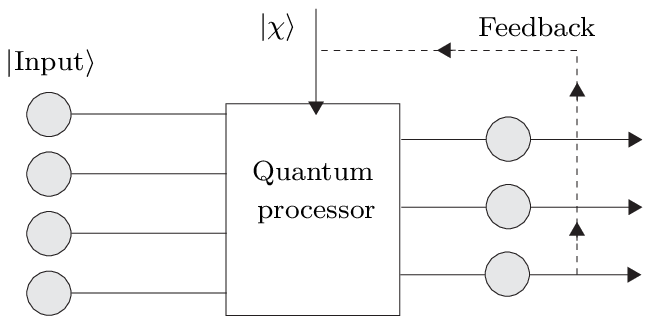}
    \label{fig:modelF}
  \end{psfrags}
  }
  \caption{Classical (a) and quantum (b) concept of learning by
    measurement and feedback. The feedback on the dynamics of the
    system is based upon the examination of the output over many
    trials. In our quantum learning scenario, quantum measurement back
    action and unitary feedback operations serve to modify the quantum
    state $|\chi\rangle$ of the coupling strengths applied in one- and
    two.-qubit gates in our quantum processor.}
\end{figure}

In this paper, we present an alternative approach to the programming
of a quantum computer based on the neural network paradigm. Neural
networks are not programmed from the beginning to accomplish a given
task, but instead their coupling parameters (equivalent to the
synaptic coupling strengths among neurons in the brain) are being
updated according to some policy -- often iteratively during a
succession of trials according to the successful or failed
performance. This adaptive learning, which draws on similarities with
the learning of human brains, in the end produces a system, rigged
with coupling parameters that enable it to solve the tasks
trained. Figure \ref{fig:modelNN} shows a model of a neural network
with a number of nodes, including input and output nodes, which are
connected with coupling strengths, which can be varied according to
impulses from external agents who check the output. The success of
neural networks as a computational concept relies on the possibility
that the resulting circuit is able to successfully solve also new
problems of the same kind as the ones trained. For a general reference
to neural networks, see \cite{neuralnetworks}. In the neural network,
illustrated in part (a) of the figure, both the coupling strengths and
the actual physical location of the information are dynamically
modified and redistributed during the learning, and one of the
strengths of neural networks is precisely the delocalization of the
memory which is believed to provide robustness against local damaging
effects.  The application of a delocalized memory for protection of
quantum information in collective local minimum energy states of
interacting many-body systems was proposed and analyzed in
\cite{PhysRevLett.98.023003}.

In the present paper, we shall focus on the iterative learning aspect
of the neural network paradigm. The training of a classical neural
network may proceed both via iterative feedback and by single step
methods. Our approach is inspired by the iterative version of the
artificial neural network paradigm. We shall investigate if
this can be implemented in a particularly simple model, where the
strengths of the coupling parameters governing a conventional set of
universal gates are treated as quantum degrees of freedom, and where a
search for optimum values of these parameters is carried out by
running the computer many times and acting back on these parameters
according to the outcome of the computation. Optimization of quantum
algorithm design has been studied, e.g., in \cite{Behrman,Navin},
where a variational method was applied to identify optimal values of
controllable parameters in a Hamiltonian to secure the optimum time
evolution of the system density matrix. The work of
\cite{Behrman,Navin} is connected to the large variety of works on
quantum control, applied in particular to femtosecond laser-chemistry
where pulse shaping is used to maximize the yield in chemical
\cite{Gerber} and biological \cite{Herek} reactions.  Our work differs
from the philosophy of \cite{Behrman,Navin} by being directed at
experimental implementation on a single quantum system. In particular,
this implies that complete knowledge of the system wave function or
density matrix is not available, and for the design of the feedback
loop one has access only to the information extracted by measurements
on a single quantum system. By the nature of quantum mechanics, this
information is random, and it is an important aspect of the analysis
that the control parameters are not only subject to adaptive changes
due to the feedback but are also modified by the measurements
themselves\footnote{For a recent review on quantum filtering and
  control see the special issue of J. Opt. B: Quant. Semiclassi. Opt
  {\bf7}, 2005, S177-S434.}. We note that pulse shaping in laser
chemistry has been successfully implemented in conjunction with
experiments, such that the molecules themselves "are responsible" for
solving the Schr\"odinger equation, and the control fields are
subsequently varied according to the experimental output by genetic
algorithms \cite{Gerber}. In this case, experiments are carried out on
large ensembles of identical systems, thus evading the issues of
stochastic measurement outcomes and quantum back action.

The use of learning strategies for quantum computers dedicated to
specials tasks, such as pattern recognition, matching of unknown
quantum states, and simulation of classical and quantum problems have
received some interest \cite{Atici,Sasaki}, and the idea of a
classical Hopfield neural network was recently combined with quantum
adiabatic computation to implement a novel quantum pattern recognition
scheme \cite{Glaser}. In \cite{Zak}, measurements and feedback
applied to both the control parameters and to subsequent input states
were discussed for the effective solution of a range of special
problems. In comparison, our work is more directed towards
optimizations of standard quantum computing, and, as exemplified below,
the optimal performance of given computational tasks.

\section{Controllable one- and two-bit gates}

If we implement the neural network with individual two-level quantum
systems taking the place of the nodes, and with access to any one- and
two-bit operations, the quantum state of the complete system is
subject to a time dependent Hamiltonian which can be parametrized with
time dependent vectors and matrices $\vec{w}_i$ and $M_{ij}$ of
coefficients multiplying operators which are, in turn, expanded on
single-qubit Pauli operators $\vec{\sigma}_i$,
\begin{align}
  H = \sum_i \vec{w}_i\cdot \vec{\sigma}_i+
  \sum_{i,j} \vec{\sigma}_i \cdot M_{ij}\vec{\sigma}_j . \label{eq:NeuralNNHamilton}
\end{align}
In the conventional approach to quantum computing based on sequential
application of one and two bit gates, the time dependence of
$\vec{w}_i$ and $M_{ij}$ is restricted to non-vanishing values
occurring only in discrete intervals.

The success probability of the computational task is a functional of
the time dependent coupling strengths $\vec{w}_i(t)$ and $M_{ij}(t)$,
and the optimal time dependent Hamiltonian may have no obvious
relation to the usual expansion on one and two-qubit gates. We imagine
that this approach can be applied to a full implementation of a
quantum computer with adjustable coupling strengths, recalling that we
deal with only a single realization of the quantum computer, and hence
the output of a single run can both be wrong for the best realization
of the quantum computer and correct for a very bad one, cf., the
finite success probabilities of the Grover and Shor
algorithms. Finding the best parameters as quickly as possible thus
belongs to the class of stochastic optimization problems, which is a
currently very active research field in applied mathematics \cite{Asmussen}.

In \cite{Bang}, a "quantum learning machine" is proposed in which
control parameters are adjusted according to a feedback mechanism
involving the record of failure and success events.  We consider in
the following a "full quantum" implementation of such a quantum
learning machine in which the $\vec{w}_i$ and $M_{ij}$ are themselves
quantum variables, implemented by coupling of our computational qubits
to auxiliary quantum systems. A model of such a neural network quantum
computer is sketched in Figure \ref{fig:modelF}. It shows the quantum
processor unit, which transforms an input state into an output state
under the action of control parameters supplied by the interaction
with the physical system represented by the state $\ket\chi$ indicated
in the figure. As in the optimal control theory problem, a
verification module acts back on the control parameter variables, and
by repeated action this system may converge and provide optimum
parameters for successful implementation of the quantum processor. In
this setup we may thus teach the computer to solve problems that are
prohibitively hard to solve but easy to verify classically. Search of
unstructured databases, satisfiability problems, factoring, etc., are
thus tasks that the successful experimentalist may train the neural
quantum computer to master.

The auxiliary quantum system may be quantum fields representing time
varying coupling strengths in a quantum manner, but as in optimal
control theory we may discretize the time or we may expand the time
dependence on a set of time dependent functions and hence represent
the coupling strengths with a finite number of degrees of freedom. The
interaction with the processing variables should not alter properties
of the quantum state $\ket\chi$ which may subsequently lead to changes
in the control parameters, and the Hamiltonian of the auxiliary system
when it is not coupled to the processor must commute with the
observables representing the coupling strengths, i.e. the coefficients
$\vec{w}_i$ and $M_{ij}$ must be Quantum Non-Demolition (QND)
observables. The logical state $\ket 1$ population in the control bit
in a two-bit control-not or a three-bit control-control-not operation
is an example of such a (discrete) implementation of $w$ and $M$
components. The auxiliary system may of course be physically
equivalent with the processing system, and e.g., part of the ions in
an ion string, using for example the approach in \cite{wang} to
implement multi-qubit gates. The idea put forth in this paper is
theoretical and conceptual, and should indeed be applicable to any
physical implementation of a quantum computer. In the following we
will discuss the idea formally and independently of any specific
physical system.

\section{Measurement back action and feedback on control parameters}

We are now ready to propose a procedure, where a circuit with register
qubits which are coupled to a number of auxiliary quantum systems is
allowed to propagate a given input. The resulting state of the
register is read out by a projective measurement, and depending on the
quality of the output, the experimental procedure consists in applying
a feedback to the control parameter components of the system, and then
repeat the computational step with the same or with other relevant
input states. After many trials, the experimentalist should have a
quantum computer at her disposal which has been trained without any
need for theoretical solutions of the time evolution problem or
knowledge about the quantum state of the control parameters produced
by the protocol.

In our proof-of-concept, we will focus in the following on the very
restricted case of a single one-dimensional control parameter with
continuous real spectrum. We thus operate with a single parameter
$\phi$ which controls a certain interaction in the quantum circuit
instead of addressing a full neural network problem with a large
number of adaptive parameters. This simple model enables the study of
the time evolution described by a parametrized family of unitary
operators $U(\phi)$.

The action of the Hamiltonian on the coupling parameter and quantum
processor product Hilbert space is given by
\begin{align}
  U_{\mathrm{tot}} = \int \d\phi \ket\phi\bra\phi \otimes U(\phi)
\end{align}
and it produces an entangled state with correlations between the
parameter eigenkets $\ket\phi$ and the result of the algorithm
$U(\phi)\ket{\symInput}$. After a projective measurement on the output
state $\ket r$, with probability $P(r)=\int \d\phi \abs{\chi(\phi)}^2
\abs{\braket{r|U(\phi)|\symInput}}^2$, the joint state becomes $\int
\d\phi \chi(\phi)\braket{r|U(\phi)|\symInput} \ket\phi  \tensorp \ket
r/\sqrt{P(r)}$.

The register is no longer entangled with the parameter-system, and the
coupling parameter wave function has been updated according to
\begin{align}
  \chi(\phi) \longrightarrow \chi(\phi) \braket{r|U(\phi)|\symInput}/\sqrt{P(r)}.  \label{eq:decimationprocess}
\end{align}
The function $\phi \mapsto \abs{\braket{r|U(\phi)|\symInput}}^2$ is
a "filter", peaked around the values of $\phi$ which produce the result $\ket r$
with high probability. Measurements of the register in state $\ket r$
hence enhance the value of the posterior wave function $\chi(\phi)$
around these values, and if we are lucky and measure only output
states $\ket r$ which pass the verification test, the quantum state
of the auxiliary system thus, by itself, converges towards the optimal
parameters. When we obtain results that do not pass the test, however,
the measurement process will reduce the probability in the optimal
regions of the parameter state, and a suitable active feedback
strategy must be applied.

We note that the optimal performance of the protocol may be obtained
for a narrow interval of the $\phi$ parameters, but although the
current wave function $\chi(\phi)$ may be peaked in this interval, we
may still obtain a negative outcome due to the non-unit success
probability of even the optimum quantum algorithm. We should hence
also make an attempt to counteract the erroneous reduction of the wave
function in regions with high success probability.

From a theorist's perspective, one may well imagine an analysis of the
states and operations involved, leading to a good feedback strategy,
but we emphasize, that we are here investigating a scheme that is
supposed to work without such extra specific knowledge. We shall
therefore only apply "natural" and quite conservative feedback
ideas. We have successfully tried push operations, where we displace
the $\phi$ argument alternatively to the left and right at every
negative outcome of the verification step, decreasing gradually the
magnitude of the push by the inverse of the square root of the number
of successful outcomes. This has the effects of smoothing out dips
stemming from negative outcomes.  Rather than stepping alternatively
to the left and right, we have also applied ideas from quantum walks
\cite{kendon-2006-364} in $\phi$-space, where we, after a measurement
with erroneous outcome, split the $\phi$ eigenkets coherently towards
both lower and higher arguments. This is for example done using a
Hamiltonian specified by $H\ket\phi = \lambda(\ket{\phi + \delta\phi}
+ \ket{\phi - \delta\phi})$, which after a time $\delta t$ leads to
the unitary feedback evolution operator
\begin{align*}
  U_{\textrm{fb}} = p_0(x)\identity  +  \sum_{l=1}^\infty p_{l}(x)\p{\T(l\delta\phi) + \T(-l\delta\phi)} ,
\end{align*}
where $x=\lambda\delta t/\hbar$, $p_{l}(x) = \sum_{j=l}^\infty (-i
x)^{2j - l}/(j!(j-l)!)$ and $\T(\Delta x)$ is the translation operator
of distance $\Delta x$. The functions $p_l(x)$ are complex, and we
have found it beneficial to counter the build-up of undesired interference
effects produced by the phases of $p_l(x)$ by applying a pseudo-random
dephasing over the parameter register.

We may design the protocol so that the initial wave
function $\chi(\phi)$ is real and almost uniform until the first
iteration outcome, which is most likely to be erroneous. The state
after this measurement thus attains a dip rather than a peak at the
optimum value of the coupling parameter, and to significantly improve
the state, we can simply apply a single step of the Grover inversion
about the mean operation on the control parameter system
\cite{grover_fast_1996}. This feedback operation thus converts the
unwanted dip into a peak at the optimum $\phi$ value. Due to generally
complex and nonuniform amplitudes, the future evolution unfortunately
does not benefit from further application of this operation, but it
gets us going in the right direction, and we henceforth proceed with
feedback actions, pushing the control parameter more gently in
response to erroneous outcomes.

\section{Numerical examples}

We will now present results of our numerical simulations of the
training of a quantum circuit to perform the Grover search algorithm
and the Shor factoring algorithm. The simulations were carried out on
a classical computer and using random number generators to simulate
the random outcome of measurements on the quantum system. We emphasize
that although the simulations proceeded by evolution of the full
quantum state, all steps in our protocol are designed to be carried
out in an experimental implementation with access only to the
sequence of right and wrong answers by the quantum processor.

\subsection{Application to Grover's search algorithm}

The generalized Grover or amplitude amplification algorithm
\cite{PhysRevLett.80.4329}, can rotate a given source state $\ket s$
close to a target state $\ket t$ using any unitary transformation $V$
in no more than $1 / \abs{V_{ts}}$ steps, where $\braket{t|V|s} =
V_{ts} \neq 0$. The algorithm uses operations that change the sign of
the $\ket s$ and $\ket t$ amplitudes \cite{grover_fast_1996}, and
using for the initial state $\ket s$ an equal superposition of all
$N_{el}$ classical computational states we get the original Grover
algorithm with $V_{ts} = 1/\sqrt{N_{el}}$.

We have in our numerical study simulated a computer which, instead of
the change of sign on the target state has been programmed to
implement an arbitrary phase-shift i.e. $\ket t \to e^{i\phi}\ket t$
instead of $\ket t \to -\ket t$, and we watch the adaptive
modification of the state $\chi(\phi)$ representing an initially
unknown phase shift towards a state with a well defined, optimum,
phase shift.

\begin{figure}
  \centering
  \includegraphics[width=0.8\textwidth]{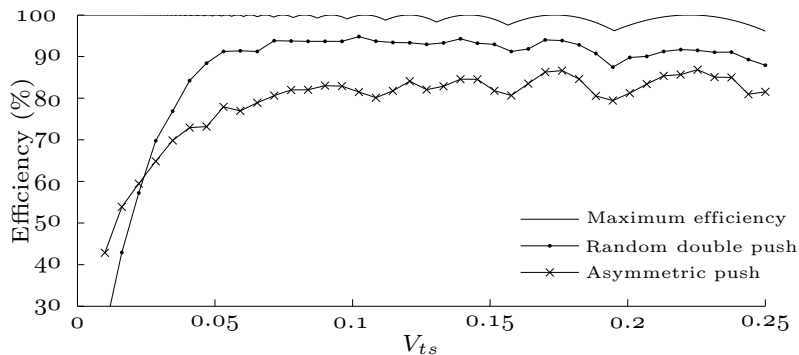}
  \caption{Average performance of Grover algorithm adaptively trained
    by single-push and double-push feedback. Results are shown after
    only \num{120} iterations of the algorithm.  The upper curve shows
    the maximal success probability of the usual Grover search
    algorithm. The size of the search space is given as $1/V_{ts}^2$, that is from \num{10000} to \num{16} elements.}
  \label{fig:oneparAlgorithmGrover}
\end{figure}

\begin{figure}
  \centering
  \includegraphics[width=0.8\textwidth]{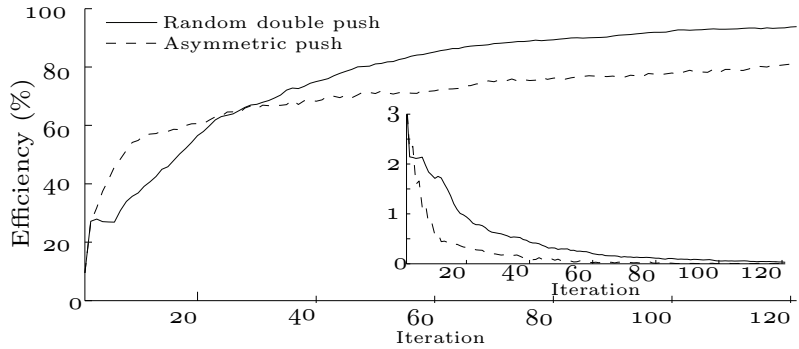}
  \caption{Average evolution of the success probability of the system
    trained with the single-push (dashed) and double-push (solid)
    feedback for $V_{ts} \approx \num{0.07}$ corresponding to the
    search of a database with $\sim$ 200 elements. Inset: The variance
    of the $\abs{\chi}^2$-distribution as a function of iteration
    number.  } \label{fig:groverTime}
\end{figure}

\begin{figure}
  \centering
  \includegraphics[width=0.8\textwidth]{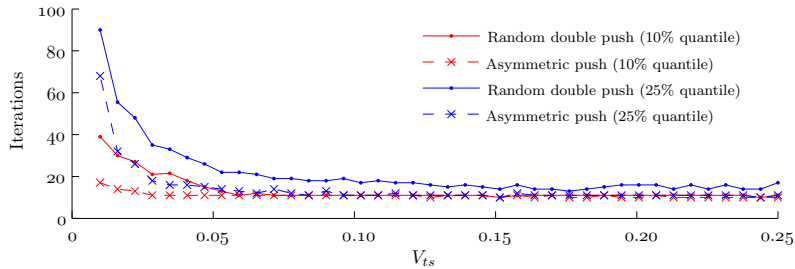}
  \caption{\SI{10}{\%} and \SI{25}{\%} quantiles of the number of
    iterations needed to obtain a probability of success of
    \SI{95}{\%} of the theoretical maximum. The data show that with
    the asymmetric push feedback, a near-optimal Grover search on
    databases with up to \num{10000} elements can be taught in less
    than 20 iterations in the best \SI{10}{\%} runs, while 10-12
    iterations suffice for smaller databases where success is achieved
    in \SI{25}{\%} of the runs.}
  \label{fig:GroverQuantile}
\end{figure}

Figure \ref{fig:oneparAlgorithmGrover} shows the efficiency of the
adaptive learning applied to the Grover algorithm, defined as the
average success-probability obtained by an ensemble of computers that
have been trained by a mere \num{120} iterations. The efficiency is
shown as a function of $V_{ts} = 1/\sqrt{N_{el}}$, corresponding to
the search of registers with a few to several thousand elements. The
lower curves in the plot refer to the simple push strategy with
decreasing magnitude and the double push, or quantum walk, strategy
followed by a randomization of the phase. The double push methods work
with high success probability in a wide interval, but both methods
degrade for large search problems with very small $V_{ts}$. For large
$V_{ts}$, i.e., for small registers, the Grover algorithm is not
always $\SI{100}{\%}$ efficient, which naturally causes the teaching
algorithms to produce worse results. The upper curve in figure
\ref{fig:oneparAlgorithmGrover} thus indicates the success probability
of the conventional Grover algorithm at different $V_{ts}$. Our
simulated curves do not reach this optimum, because they represent an
average, including contributions from entirely unsuccessful teaching
attempts. Such unsuccessful attempts are the ones, where no tendency
of improved success is observed in the measurement read-out, and such
runs of the learning algorithm would typically be discarded in an
experimental implementation of the scheme.

In Fig.\ref{fig:groverTime} the quality of the algorithm is plotted as
a function of the number of iterations, and the insert shows the
variance of the parameter $\phi$ according to the distributions
$|\chi(\phi)|^2$ found in several runs of the protocol. After
\num{100} iterations the learning saturates and any further increase
in quality becomes more or less negligible.\par

Note that the variance of $\abs{\chi(\phi)}^2$ is not a quantity that
we assume available, or for that sake, of relevance in an
implementation of our proposal. The plot only shows that $\phi$
localizes with the application of our feedback scheme.

A different representation of the simulation data can be seen in
figure \ref{fig:GroverQuantile}, where we show the required number of
iterations where at least \SI{10}{\%} (\SI{25}{\%}) of our simulations
have led to a performance better than \SI{95}{\%} of the theoretical
maximum. Thus for a wide range of register sites only about
\num{10}-\num{20} iterations are needed for the \SI{25}{\%} best runs to reach \SI{95}{\%} of the maximum success probability.

\subsection{Application to Shor's factoring algorithm}

We have also applied the learning algorithm to the discrete Fourier
transform which is an essential part of the Shor factoring
algorithm. The discrete Fourier transform implements controlled phase
shifts $\phi_m=\pi/2^m$, falling off with the separation $m$ between
the qubit register positions in a binary representation of
integers. It has been proposed, in order to gain speed and to reduce
the effects of decoherence and noise \cite{PhysRevA.54.139}, to
implement an approximate Fourier transform, which only considers
couplings up to a given maximum distance. If the gates are thus
truncated, one may well speculate that other phases than the usual
choice leads to better performance.  This is verified by numerical
computations showing an efficiency-gain of up to \SI{12}{\%} depending
on the degree of approximation. For example a quantum Fourier
transform of \num{14} qubits with nearest neighbor couplings could be
improved \SI{12.3}{\%} just by chosing apropriate phases. See Table
\ref{tab:improvement} for more examples. Although the identification
of optimum phases seems to converge well on not too large systems, the
problem constitutes a good example of our aims to find optimum
parameters in a quantum processor by measurements and feedback. We have
thus treated the unknown phase(s) as provided by an auxiliary quantum
system and applied the same feedback algorithm as described above in
order to teach a Fourier transform network with freedom in the choice
of phases to solve its task optimally.

\begin{table}
  \centering
\begin{tabular}{rr|@{\hspace{0.5cm}}r@{\hspace{0.5cm}}r@{\hspace{0.5cm}}r}
 &  & \multicolumn{3}{c}{Separation, $m$} \\
 &  & 1 & 2 & 3 \\
\hline
 \multirow{4}{*}{\begin{sideways}Qubits\end{sideways}} & 6 & \num{4.5} &  \num{1.3} &    \\
  & 8 & \num{7.3} &  \num{3.2} &  \num{0.5}  \\
  & 10 & \num{10.5} &  \num{6.0} &  \num{1.2}  \\
  & 12 & \num{11.7} &  \num{7.5} &  \num{2.1}  \\
  & 14 & \num{12.3} &  \num{11.4} &  \num{3.0} \\
  \end{tabular}
  \caption{Improvement in percent of approximate quantum Fourier transform by use of non-standard phases on registers with different numbers of qubits and with truncation of two-qubit gates at separations 1, 2 and 3.}
  \label{tab:improvement}
\end{table}

\begin{figure}
  \centering
\noindent
\subfigure[~ \label{fig:AQFT2FeedbackQuality}]{  \includegraphics[height=0.28\textheight]{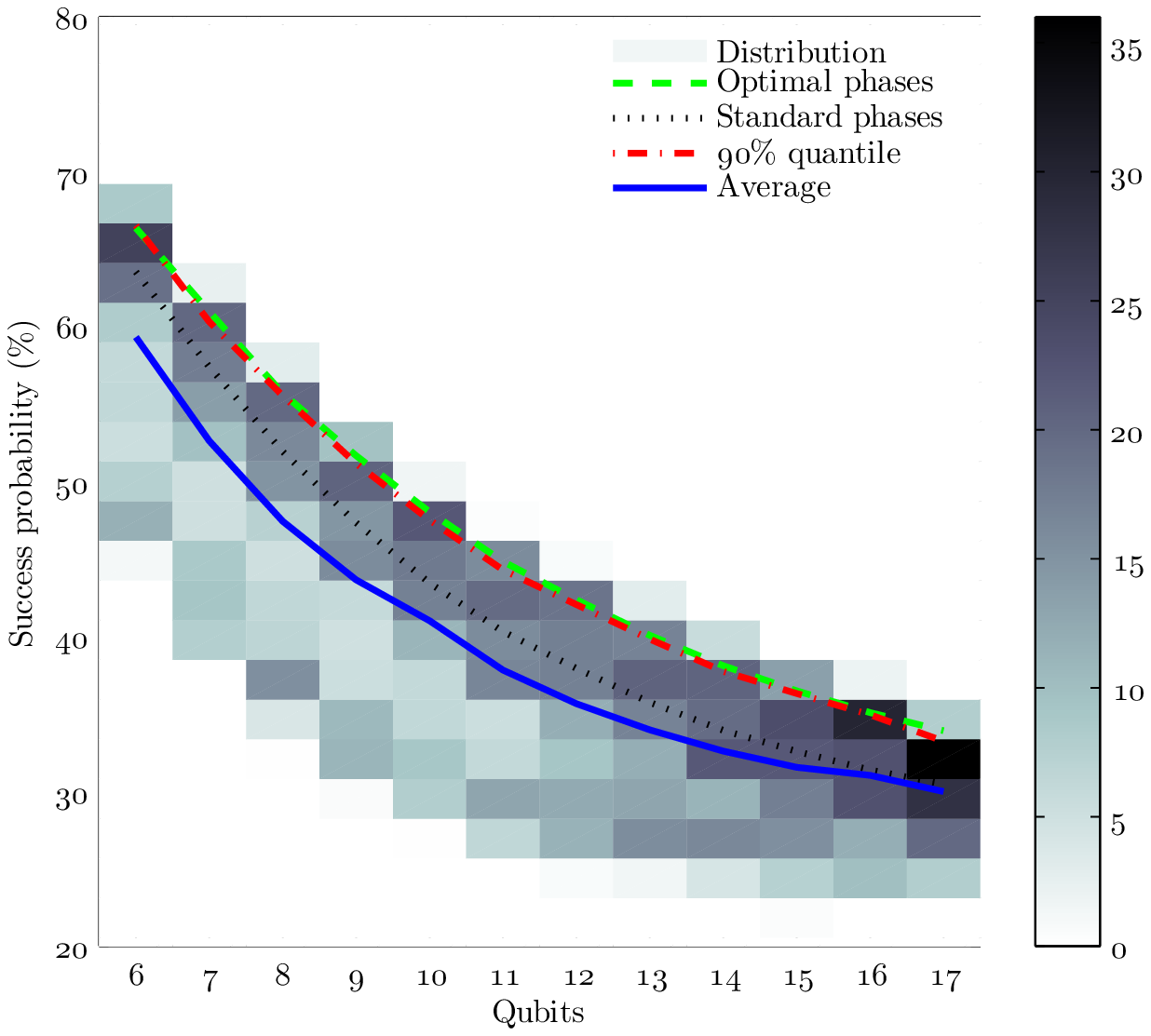}}
\subfigure[~ \label{fig:AQFT3FeedbackQuality}]{  \includegraphics[height=0.28\textheight]{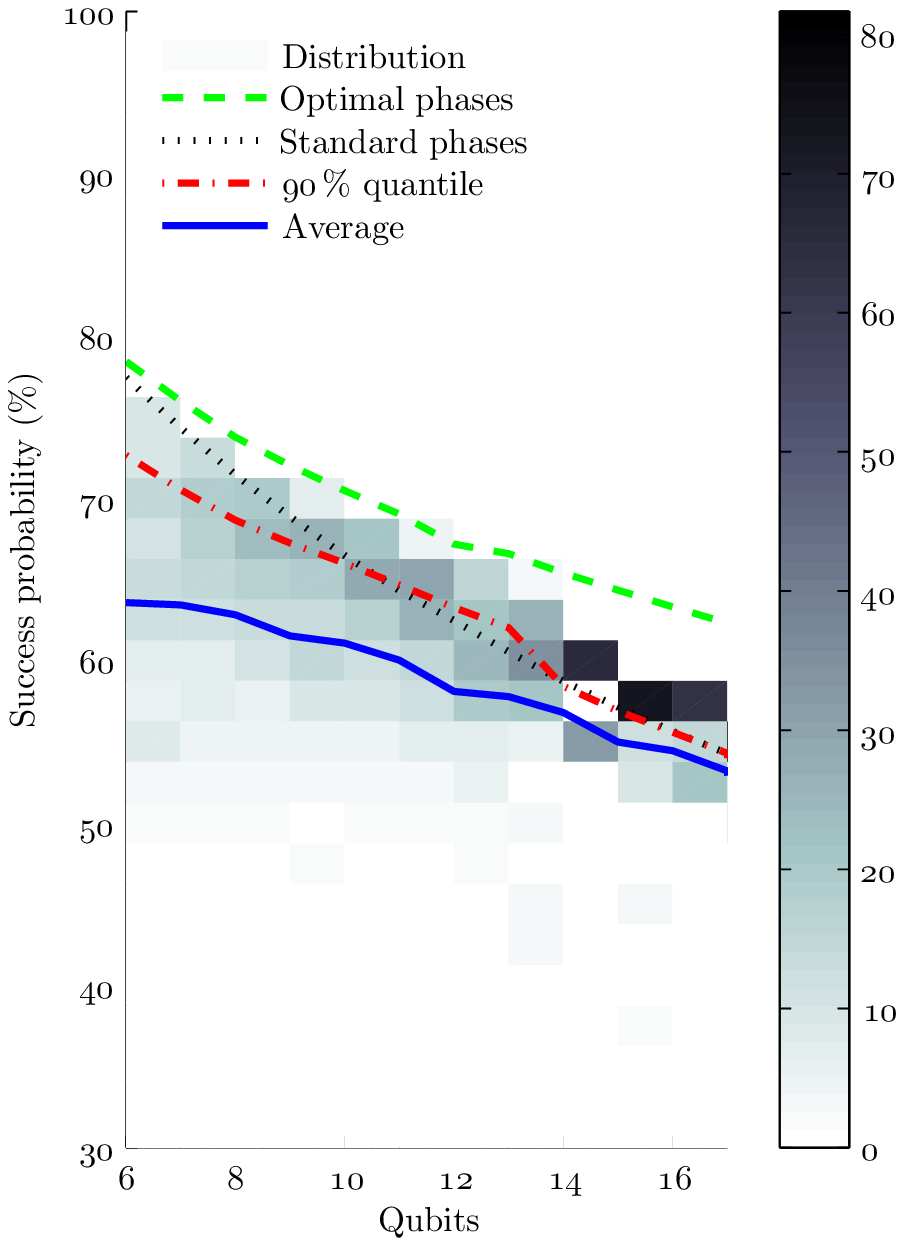}}
\caption{Distribution of the success-probability of
  the test-runs (a) for nearest neighbor couplings and (b) for nearest
  and next-nearest neighbor couplings. The shading shows how large a
  fraction of our test runs have success probability within different
  2.5 \% wide percent intervals after 120 iterations for 6 to 17
  qubits. The solid curve indicates the average success probability
  and the dash-dotted curve shows the 90\% quantile. The dotted and
  dashed line shows the average success probability for standard and
  optimal phases respectively. }
 
\end{figure}

Fig. \ref{fig:AQFT2FeedbackQuality} shows the results of using the
asymmetric push algorithm on the approximate quantum Fourier transform
with nearest neighbor coupling ($m = 1$) as described in
\cite{PhysRevA.54.139}. After \num{120} iterations on a number of
independent trials, we end up with different states of the control
parameter and, hence, with different success probabilities in the
subsequent performance of the computation. The shading in the figure
indicates for different sizes of the register between 6 and 17 qubits
the fraction of events with success probabilities in 2.5 \% wide
intervals. For comparison, the solid curve in the plot shows the
average efficiency, while the dotted line indicates the average success
probability using standard phases.

The 90 \% quantile (the dash-dotted line) shows the success
probability delimiting the 10 \% best from the 90 \% worst
performances and show that a significant fraction of runs result in
very good performance. These particularly successful runs can, to some
extent, be identified by the experimenter through a larger number of
successful outcomes of the verification in the 120 iterations. The
dashed line in the figure shows the theoretical maximum success
probability, which we calculated by using standard optimization
algorithms. As shown in the figure, the 90\% quantile lies very close
to this maximum, indicating that 10\% of the parameters found by our neural learning approach are
very close to optimal.

We have also attempted to optimize the $m = 2$-case, i.e. a problem
with two unknown parameters, and an algorithm producing viable
solutions was found, but the probability of success was not as
illustrative as the cases presented for the $m = 1$ case (see figure \ref{fig:AQFT3FeedbackQuality}) . This we ascribe to the
higher dimensionality of the problem, and it is clearly a challenge
for any feedback strategy to apply appropriate corrections to
multi-dimensional control parameters.

\section{Conclusion}

In conclusion, we have proposed to treat coupling strengths in a
quantum circuit as quantum variables and to apply feedback strategies
on these variables to adaptively teach the circuit to solve given
tasks. The proposal is formulated such that it may be implemented in a
suitable experimental set-up, and such that there is no need for an
elaborate parallel theoretical calculation. One- and two-parameter
simulations confirm the viability of the proposal, but it should be
recalled that with the extension to many coupling parameters, the
optimum feedback in the corresponding multi-dimensional parameter
space is highly non-trivial. In the latter case, solutions may be
found which rely on superposition states or perhaps entangled states
of the coupling parameter systems, and in such cases the algorithm
optimization goes far beyond optimum classical solutions for the
control parameters in Eq. (\ref{eq:NeuralNNHamilton}). It should be
emphasized that the applied feedback strategies have been quite simple
and general, and that the resulting performance is both remarkable and
encouraging. For physical implementation, we note that the coupling
parameter variables can be incorporated on equal footing with the
register qubits of the quantum computer, but they should be restricted
to QND behavior, and they should be easy to address by the feedback.
In a longer perspective our proposal may open the possibility to
identify optimum devices for few bit operations such as operations
within error correcting codes \cite{Steane,Shor} and for the training
of a many qubit "quantum brain", that accomplishes very difficult
tasks.


\begin{thebibliography}{10}

\bibitem{NielsenChuang}
M.~A. Nielsen and I.~L. Chuang.
\newblock {\em Quantum Computation and Quantum Information}.
\newblock Cambridge University Press, 2000.

\bibitem{neuralnetworks}
J.~Hertz, R.~G. Palmer, and A.~Krogh.
\newblock {\em Introduction to the Theory of Neural Computation}.
\newblock Addison Wesley, 1991.

\bibitem{PhysRevLett.98.023003}
M.~Pons, V.~Ahufinger, C.~Wunderlich, A.~Sanpera, S.~Braungardt, A.~Sen(De),
  U.~Sen, and M.~Lewenstein.
\newblock Trapped ion chain as a neural network: Error resistant quantum
  computation.
\newblock {\em Phys. Rev. Lett.}, 98(2):023003, Jan 2007.

\bibitem{Behrman}
E.~C. Behrman, J.~E. Steck, P.~Kumar, and K.~A. Walsh.
\newblock Quantum algorithm design using dynamic learning.
\newblock {\em Quantum Information and Computation}, 8:12, 2008.

\bibitem{Navin}
N.~Khaneja, R.~Brockett, and S.0~J. Glaser.
\newblock Time optimal control in spin systems.
\newblock {\em Phys. Rev. A}, 63(3):032308, Feb 2001.

\bibitem{Gerber}
T.~Brixner, N.~H. Damrauer, P.~Niklaus, and G.~Gerber.
\newblock Photoselective adaptive femtosecond quantum control in the liquid
  phase.
\newblock {\em Nature}, 414:57--60, November 2001.

\bibitem{Herek}
J.~L. Herek, W.~Wohlleben, R.~J. Cogdell, D.~Zeidler, and M.~Motzkus.
\newblock Quantum control of energy flow in light harvesting.
\newblock {\em Nature}, 417:533--535, May 2002.

\bibitem{Atici}
A.~Atici and R.A. Servedio.
\newblock {Improved Bounds on Quantum Learning Algorithms}.
\newblock {\em Quantum Information Processing}, 4(5):355--386, 2005.

\bibitem{Sasaki}
M.~Sasaki and A.~Carlini.
\newblock {Quantum learning and universal quantum matching machine}.
\newblock {\em Phys. Rev. A}, 66(2):22303, 2002.

\bibitem{Glaser}
R.~Neigovzen, J.~Neves, R.~Sollacher, and S.~J. Glaser.
\newblock Quantum pattern recognition with liquid state nmr.
\newblock {\em arxiv: 0802.1592}, February 2008.

\bibitem{Zak}
M.~Zak.
\newblock {Quantum Analog Computing}.
\newblock {\em Chaos, Solitons and Fractals}, 10(10):1583--1620, 1999.

\bibitem{Asmussen}
S.~Asmussen and P.W. Glynn.
\newblock {\em {Stochastic simulation: algorithms and analysis}}.
\newblock Springer, New York, 2007.

\bibitem{Bang}
J.~Bang, J.~Lim, MS~Kim, and J.~Lee.
\newblock {Quantum Learning Machine}.
\newblock {\em arXiv:0803.2976}, 2008.

\bibitem{wang}
X.~Wang, A.~S{\o}rensen, and K.~M{\o}lmer.
\newblock Multibit gates for quantum computing.
\newblock {\em Phys. Rev. Lett.}, 86:3907, April 2001.

\bibitem{kendon-2006-364}
V.~M. Kendon.
\newblock A random walk approach to quantum algorithms.
\newblock {\em Phil Trans Roy Soc A}, 364(1849):3407--3422, Dec 2006.

\bibitem{grover_fast_1996}
L.~K. Grover.
\newblock A fast quantum mechanical algorithm for database search.
\newblock In {\em STOC '96: Proc. 28. Ann. ACM symp on theory of computing},
  pages 212--219, New York, USA, 1996. ACM.

\bibitem{PhysRevLett.80.4329}
L.~K. Grover.
\newblock Quantum computers can search rapidly by using almost any
  transformation.
\newblock {\em Phys. Rev. Lett.}, 80(19):4329--4332, May 1998.

\bibitem{PhysRevA.54.139}
A.~Barenco, A.~Ekert, K~Suominen, and P.~T\"orm\"a.
\newblock Approximate quantum fourier transform and decoherence.
\newblock {\em Phys. Rev. A}, 54(1):139--146, Jul 1996.

\bibitem{Steane}
A.~M. Steane.
\newblock Error correcting codes in quantum theory.
\newblock {\em Phys. Rev. Lett.}, 77(5):793--797, Jul 1996.

\bibitem{Shor}
P.~Shor.
\newblock Scheme for reducing decoherence in quantum computer memory.
\newblock {\em Phys. Rev. A}, 52(4):R2493--R2496, Oct 1995.

\end{thebibliography}
\end{document}